\documentclass[10pt]{article}
\usepackage{amsfonts}

\begin{document}

\title{The Color-Flavor Transformation and a New Approach to
       Quantum Chaotic Maps}

\author{%
  M R Zirnbauer%
  \footnote{%
    Institut f\"{u}r Theoretische Physik, Universit\"{a}t zu K\"{o}ln,
    Germany.
    email: \texttt{zirn@thp.Uni-Koeln.de}.
  }
}

\date{}
\maketitle

\abstract{The color-flavor transformation is a mathematical result
  that has applications to problems as diverse as lattice gauge
  theory, random network models, and dynamical systems.  Several
  variants are described, and an outline of the proof is given.  It is
  then shown how to use the transformation to set up a field theoretic
  formalism for quantum chaotic maps.}

\section{Introduction}

Building on the supersymmetric field theory of disordered conductors
and the lessons learned from it, Andreev et al. \cite{a3s} have
recently proposed a fresh and, I think, exciting approach to quantum
Hamiltonian systems the classical dynamics of which is chaotic (see
the contribution of B.D. Simons to these proceedings).  To introduce
the subject of my talk, I wish to hint at another intriguing analogy,
which links quantum chaotic and disordered systems with the low energy
chiral sector of quantum chromodynamics (QCD).

Let me start by recalling that the low energy degrees of freedom of
QCD are particle-hole excitations, namely quark-antiquark states that
are called mesons (or pions).  The effective field theory governing
the mesons and the interactions between them is known to be a
nonlinear $\sigma$ model.  Similarly, the long wave length degrees of
freedom of a disordered conductor are particle-hole like excitations,
namely diffusion modes, whose interactions are described again by a
field theory of the nonlinear $\sigma$ model type.  This analogy at
the superficial or phenomenological level has a firm basis in the
modeling of the systems.  Indeed, in the strong coupling limit of
lattice QCD the nonabelian gauge field seen by the quarks is
represented by matrices placed on the links of the lattice and drawn
at random from the gauge group ${\rm SU}(N)$.  Similarly, in models of
two-dimensional electrons in a strong magnetic field, random unitary
matrices modeling the disorder potential are assigned to the links of
a directed network.

A mathematical physicist will then ask how the reduction from random
unitary matrices on links to a nonlinear $\sigma$ model is achieved.
The first point of this talk is to report the recent discovery of a
reduction scheme which is {\it nonperturbative} and circumvents the
traditional use of diagrammatic techniques to resum the leading terms
in a perturbation expansion.  The new approach is based on an identity
that transforms a certain class of integrals over the gauge group
${\rm U}(N)$ into integrals over the corresponding mesonic or
``flavor'' degrees of freedom.  Taking the terminology from QCD, I
call this identity the {\it color-flavor transformation}.

Because of the associated gain in efficiency, the color-flavor
transformation most naturally applies to cases such as QCD in the
large-$N$ limit, where the number of colors is big and the number of
flavors small.  However, the transformation promises nontrivial
applications also in other cases.  It offers, in particular, the
possibility of constructing a supersymmetric field theory of quantized
symplectic maps.  More precisely, it allows to express ${\rm U}(1)$
phase averaged products of the Green's functions of such a map as
integrals over nonlinear $\sigma$ model fields.  The latter
presentation appears particularly well suited for extracting the
universal behavior that is expected on the basis of the
Bohigas-Giannoni-Schmit conjecture \cite{bgs}, stating that the
spectral statistics of quantum chaotic systems in the universal limit
is Wigner-Dyson.

\section{Color-Flavor Transformation}

I'll start with a simple equation that motivates and illustrates the
general form of the transformation.  If $\psi_0, \psi_1$ are complex
fermionic sources, the following statement is immediate:
        \[
        \int_0^{2\pi} {\rm d}a \, \exp \left( \bar\psi_{1} e^{ia} \psi_{0}
         + \bar\psi_{0} e^{-i{a}} \psi_{1} \right)
        = 2\pi \left( 1 + \bar\psi_{1}\psi_{0}\bar\psi_{0}\psi_{1} \right)
        = \pi \int_{\mathbb C} {{\rm d}z {\rm d}{\bar z}
          \over (1+{\bar z}z)^3} \exp \left( z \bar\psi_{1} \psi_{1}
          -\bar z \bar\psi_{0} \psi_{0} \right) \;.
        \]
The first equality sign does the integral over $a$ (the ``random gauge
field''), and the second organizes the quartic term into bilinears
$(\bar \psi_1 \psi_0 \bar \psi_0 \psi_1 = -\bar\psi_1\psi_1 \times
\bar\psi_0\psi_0)$, which are sent back to the exponent by integrating
over the complex numbers.

Why isn't $\psi_0$ paired with $\psi_1$, and $\bar\psi_0$ with $\bar
\psi_1$?  The answer is that in the form given, the scheme permits the
following remarkable generalization:
        \[
        \int_{{\rm U}(N)} dU \,
        \exp \left( \bar\psi_{1\mu}^i U^{ij} \psi_{0\mu}^j 
        + \bar\psi_{0\nu}^j \bar U^{ij} \psi_{1\nu}^i \right)
        = \int_{{\mathbb C}^{n\times n}} d\mu_N(Z,{\bar Z}) \,
        \exp \left( \bar\psi_{1\mu}^i Z_{\mu\nu}^{\vphantom{i}} \psi_{1\nu}^i
        -{\bar\psi}_{0\nu}^j {\bar Z}_{\mu\nu}^{\vphantom{i}} \psi_{0\mu}^j 
        \right) \;,
        \]
which deals with the case of $n$ ``flavors'' ($\mu,\nu=1,...,n$), and
$N$ ``colors'' ($i,j=1,...,N$) transforming according to the vector
representation of the gauge group ${\rm U}(N)$.  The integration
measure $dU$ on the left-hand side is the Haar measure of ${\rm
  U}(N)$.  Summation over repeated indices is always assumed.  The
integral on the right-hand side is over all complex $n\times n$
matrices $Z$, with adjoint ${Z^\dagger}_{\nu\mu} = \bar Z_{\mu\nu}$,
and
        \[
        d\mu_N(Z,\bar Z) = {\rm const} \times 
        {\rm Det}(1+Z^\dagger Z)^{-2n-N} \prod_{\mu,\nu=1}^{n} 
        {\rm d}Z_{\mu\nu} {\rm d}{\bar Z}_{\mu\nu} \;.
        \]
The geometric content behind this expression is that the matrix $Z$
parameterizes the compact symmetric space $M = {\rm U}(2n)/{\rm U}(n)
\times {\rm U}(n)$, and the ${\rm U}(2n)$-invariant volume form on $M$
pulls back to the integration measure $d\mu_0(Z,\bar Z)$.

Note that the above formula is justifiably called the {\it
  color-flavor transformation}.  On the left-hand side the flavor
degrees of freedom are uncoupled, and color interacts through the
gauge group element $U \in {\rm U}(N)$.  On the right-hand side, the
roles have been reversed: now it is the color degrees of freedom that
enter multiplicatively, while flavor interacts by the ``mesonic
field'' $Z$.

The integral on the left-hand side is of the type encountered in
lattice gauge theory with dynamical fermions.  I claim that by using
the color-flavor transformation one can in fact verify that large-$N$
quantum chromodynamics in the strong coupling limit is equivalent to a
weakly coupled theory of mesons \cite{witten}.  Whether the scheme can
be adapted to deal with the weak coupling (or asymptotically free)
limit of QCD is under investigation.

The above formula is just one out of a large number of variants:

\begin{enumerate}
\item A similar formula holds for {\it bosonic} sources $\phi$ instead
  of the fermionic $\psi$.  To get the bosonic version, one simply
  makes the formal replacements $N \to -N$ and $\bar Z \to -\bar Z$,
  and restricts the domain of integration by requiring $1-Z^\dagger Z
  > 0$.  [These steps imply that $M$ is replaced by its noncompact
  analog ${\rm U}(n,n)/{\rm U}(n)\times {\rm U}(n)$.]  Convergence of
  all integrals now requires $N \ge 2n$.
\item In Ref.~\cite{mrz_circular} a supersymmetric version with $n$
  bosonic (B) and $n$ fermionic (F) flavors was proved.  More
  generally, the color-flavor transformation exists also for $n_{\rm
    B} \not= n_{\rm F}$ bosonic and fermionic flavors, as long as $N
  \ge 2(n_{\rm B}-n_{\rm F})$.
\item The gauge group ${\rm U}(N)$ can be replaced by any one of the
  compact Lie groups ${\rm O}(N)$ or ${\rm Sp}(N)$.  For ${\rm Sp}(N)$
  a supersymmetric version of the transformation formula was given in
  Ref.~\cite{mrz_circular}, and the case ${\rm SO}(N)$ will be
  discussed below.
\item One can also pass from groups to symmetric spaces, that is to
  say a color-flavor transformation exists for all random matrix
  models based on one of the large families of Riemannian symmetric
  spaces of the compact type.  As an example, consider the set of
  symmetric unitary matrices $S$, which is isomorphic to ${\rm
    U}(N)/{\rm O}(N)$. The trick that works in this case is to use the
  canonical projection ${\rm U}(N) \to {\rm U}(N)/{\rm O}(N)$ by $U
  \mapsto UU^{\rm T} =: S$, in conjunction with the Gaussian integral
        \[
        \exp \left( \bar\psi^i S^{ij} \psi^j \right) = 
        \int \exp\left( - \bar\Psi^k \Psi^k +
        \bar\psi^i U^{ik}\Psi^k + \bar\Psi^k U^{jk}\psi^j \right) \;.
        \]
Since the invariant measures match under projection, this relates the 
integral over $S$ to an integral over $U \in {\rm U}(N)$, with the
number of flavors doubled.
\end{enumerate}

\section{Outline of Proof}

I'll sketch the general strategy at the example of bosonic sources
$\phi_\mu^i$ in the fundamental representation of the gauge group
${\rm SO}(N)$.  In this case, the claim is
        \[
        \int_{{\rm SO}(N)} dO \, 
        \exp \left( \bar\phi_\mu^i O^{ij} \phi_\mu^j \right)
        = \int_D d\mu_N(Z,\bar Z) \, 
        \exp\left( \bar\phi^i_{\mu} Z_{\mu\nu}^{\vphantom{i}}
        \bar\phi^i_{\nu} + {\rm c.c.} \right) \;.
        \]
The mesonic integral here runs over the set $D$ of all complex {\it 
  symmetric} $n \times n$ matrices $Z$ with $1-\bar ZZ > 0$, and the
integration measure is
        \[
        d\mu_N(Z,\bar Z) = {\rm const} \times 
        {\rm Det}(1-\bar ZZ)^{N-n-1} \prod_{\mu\le\nu} 
        {\rm d}Z_{\mu\nu} {\rm d}{\bar Z}_{\mu\nu} \;.
        \]
The domain $D$ is diffeomorphic to the noncompact symmetric space
${\rm Sp}(n,{\mathbb R}) /{\rm U}(n)$.

For the purpose of proving the present version of the color-flavor
transformation, it is convenient to quantize the classical sources:
$\phi_\mu^i\to b_\mu^i$, $\bar\phi_\mu^i \to \bar b_\mu^i$, where
$b,\bar b$ are operators that satisfy the canonical boson commutation
relations $[b_\mu^i,\bar b_\nu^j] = \delta^{ij} \delta_{\mu\nu}$ and
act in a Fock space with vacuum $b_\mu^i | 0 \rangle = 0$.  The set of
all single-boson operators $b b$, ${\bar b}{\bar b}$, and $\bar b b +
b\bar b$ (indices omitted), define a representation of the symplectic
Lie algebra ${\rm sp}(nN,{\mathbb C})$ on Fock space.  Two prominent
subalgebras are ${\rm sp}(n,{\mathbb C})$ generated by the color
singlet operators ${b}_\mu^{i} {b}_\nu^{i}$, ${\bar b}_\mu^{i} {\bar
  b}_\nu^{i}$ and ${\bar b}_\mu^{i} b_\nu^i + b_\nu^i {\bar
  b}_\mu^{i}$, and ${\rm so}(N,{\mathbb C})$ generated by the flavor
singlet operators ${\bar b}_\mu^i b_\mu^j - {\bar b}_\mu^j b_\mu^i$.
The latter subalgebra is the maximal subalgebra in ${\rm
  sp}(nN,{\mathbb C})$ that commutes with the former, and vice versa.

Now we consider two types of coherent state:
        \[
        |\phi\rangle = \exp \left( \phi_\mu^{i} {\bar b}_\mu^{i} \right) 
        | 0 \rangle \;, \quad {\rm and} \quad
        | Z \rangle = \exp \left( {\bar b}_\mu^{i} 
        Z_{\mu\nu}^{\vphantom{i}} {\bar b}_\nu^{i} \right) | 0 \rangle \;.
        \]
Their overlap is 
        \[
        \langle \phi | {Z} \rangle = \exp \left( \bar\phi_\mu^i 
        Z_{\mu\nu}^{\vphantom{i}} \bar\phi_\nu^{i} \right) \;.
        \]
By using standard results from Lie group theory, one shows that the
operator $P$ defined by
        \[
        P = \int_D d\mu_N(Z,\bar Z) \, | Z \rangle \langle Z | 
        \]
projects Fock space onto the unique subspace that contains the vacuum 
and carries an irreducible unitary group action $O \mapsto T_O := \exp
\left( {\bar b}_\mu^i (\ln O)^{ij} b_\mu^j \right)$ of ${\rm SO}(N)$.
An alternative implementation of $P$ is by the Haar integral
        \[
        P = \int_{{\rm SO}(N)} dO \, T_O \;.
        \]
With these tools in hand, the color-flavor transformation is simply
proved by the following computation:
        \begin{eqnarray}
        &&\int_D d\mu_N(Z,\bar Z) \, \exp\left(\bar\phi_\mu^i 
        Z_{\mu\nu}^{\vphantom{i}} \bar\phi_\nu^i + {\rm c.c.} \right)
        = \int_D d\mu_N(Z,\bar Z) \, \langle \phi | Z \rangle
        \langle Z | \phi \rangle \nonumber \\
        &&= \langle \phi | P | \phi \rangle
        = \int\limits_{{\rm SO}(N)} dO \ 
        \langle \phi | T_O | \phi \rangle
        = \int\limits_{{\rm SO}(N)} dO \ 
        \exp \left( {\bar\phi}_\mu^i O^{ij} \phi_\mu^j \right) \;. 
        \nonumber\end{eqnarray}

\section{Application to Quantum Maps}

On a symplectic manifold $(M,\omega)$ of dimension $2d$ consider a
symplectic map $\chi : M \to M$.  Quantization will turn $\chi$ into a
unitary operator $U$ on a Hilbert space ${\cal H}_N$ of finite
dimension $N = (2\pi\hbar)^{-d} \int_{M}\omega^{\wedge d}$.  It is
then interesting to ask what one can say about the correlations
between the eigenvalues $e^{i\theta_\mu}$ $(\mu \in \{1,...,N\})$ of
the quantum ``map'' $U$ when the classical map $\chi$ is chaotic.

An informative statistic is the {\it pair correlation function} $C$.
Its value on a testfunction $f : {\rm U}(1) \to {\mathbb R}$ is
defined to be
        \[
        C[f] = {2\pi\over N^2} \sum_{\mu,\nu=1}^N
        f ( e^{i(\theta_\mu - \theta_\nu)} ) - 
        \int\limits_0^{2\pi} f(e^{i\theta}) {\rm d}\theta \;.
        \]
Given the Fourier expansion $f(e^{i\theta}) = (2\pi)^{-1} \sum_{l\in
  {\mathbb Z}} {\tilde f}_l \; e^{il\theta}$, the pair correlation
function can be written in the equivalent form
        \[
        C[f] = {2\over N^2} \sum_{l=1}^\infty {\rm Re}{\tilde f}_l \; 
        \big| {\rm Tr} U^l \big|^2 \;.
        \]
In the physics literature one usually considers a related object, the 
so-called two-level correlation function $R_2$, which is obtained by
evaluating $C$ on the $\delta$-distribution $\delta_r = (2\pi)^{-1}
\sum_{l\in{\mathbb Z}} e^{il(\theta-r)}$ centered at $r = 2\pi s/N$:
        \[
        R_2(s) = C[\delta_{2\pi s/N}] = {2\over N^2} \sum_{l=1}^\infty
        \cos(2\pi s l/N) \; \big| {\rm Tr} U^l \big|^2 \;.
        \]
For chaotic systems, very little is known about $R_2(s)$ (and other
correlation functions) from a rigorous standpoint.  There exists,
however, a conjecture by Bohigas, Giannoni and Schmit \cite{bgs},
stating that the correlation functions for $N \to \infty$ tend to a
universal limit given by random-matrix theory.  In particular, the
two-level correlation function of chaotic systems with broken
time-reversal symmetry is predicted to be
        \[
        R_2(s) = \delta(s) - {\sin^2 \pi s \over \pi^2 s^2} \;.
        \]
It is a challenging and longstanding problem in mathematical physics
to specify the precise conditions on the quantum map $U$ under which
this conjecture holds true.

How does the color-flavor transformation come in here?  To that end,
consider the generating function
        \[
        \Omega_U(\alpha_+,\beta_+;\alpha_-,\beta_-)
        = \frac{{\rm Det}(1-\beta_+U)\,{\rm Det}(1-\beta_-U^\dagger)}
        {{\rm Det}(1-\alpha_+U)\,{\rm Det}(1-\alpha_-U^\dagger)} \,,
        \]
where $\alpha_\pm, \beta_\pm$ are complex numbers, and take an average
over the unit circle ${\rm S}^1$ in ${\mathbb C}$:
        \[
        \hat\Omega_U(...) = 
        \int_0^{2\pi}{d\varphi\over 2\pi}\, \Omega_{\exp(i\varphi)U}(...)\;.
        \]
This average over the spectral domain ${\rm S}^1 \simeq {\rm U}(1)$ 
of the unitary operator $U$ corresponds to the energy averaging
employed in the work of Andreev et al. \cite{a3s} on Hamiltonian
systems.  It is an easy computation to show that the two-level
correlation function can be extracted from $\hat\Omega_U$ by
        \[
        R_2(s) = {1\over 2N^2} {\rm Re} {\partial^2 \over \partial\theta
        \partial\theta'} \hat\Omega_U ( e^{i\theta+2\pi is/N},
        e^{-i\theta+2\pi is/N} ; e^{-i\theta'}, e^{i\theta'} ) 
        \Big|_{\theta = \theta' = 0} \;.
        \]

The starting point for setting up a field theoretic formalism for
quantum maps is the following formula:
        \[
        \Omega_U(\gamma_+;\gamma_-) = \int
        \exp \left( - \bar\Psi_+ ( 1 - \gamma_+ \otimes U ) \Psi_+
        - \bar\Psi_- ( 1 - \gamma_- \otimes U^\dagger ) \Psi_- \right) \;,
        \]
which expresses $\Omega_U$ as a Gaussian integral over ``supervectors''
$\Psi_\pm$.  Here $\gamma_\pm = {\rm diag}(\alpha_\pm,\beta_\pm)$ acts
in a ${\mathbb Z}_2$-graded space ${\mathbb C}^{1|1}$, and $\Psi_\pm$
are elements of what Berezin calls the Grassmann envelope of ${\mathbb
  C}^{1|1} \otimes {\cal H}_N$.

In the next step we do the phase average $\Omega_U \to \hat\Omega_U$,
by applying the color-flavor transformation with $n = 1$ ``colors''
and $N = {\rm dim}{\cal H}_N$ ``flavors''.  (Note that we have
switched notation $N \leftrightarrow n$ to emphasize the inequality $N
\gg n$ in the present application of the transformation formula.)
Then, on carrying out the Gaussian superintegral over $\Psi,\bar\Psi$
we get
        \begin{equation}
        \hat\Omega_U (\gamma_+ ; \gamma_- ) =
        \int_{D_N} D\mu_N(Z,\tilde Z) \, {\rm SDet}(1-\tilde ZZ) 
        \, {\rm SDet} \left( 1 - \tilde Z (\gamma_+ \otimes U) Z 
        (\gamma_- \otimes U^\dagger) \right)^{-1} \;,
        \label{Omega}
        \end{equation}
where ${\rm SDet}$ is the superdeterminant.  The integration variables
$Z,\tilde Z$ are complex supermatrices,
        \[
        Z = \pmatrix{Z^{\rm BB} &Z^{\rm BF}\cr 
                     Z^{\rm FB} &Z^{\rm FF}\cr}
        = \sum_{\mu,\nu=1}^N \, \sum_{\sigma,\tau={\rm B,F}} 
        Z_{\mu\nu}^{\sigma\tau} \, E_{\sigma\tau}\otimes E_{\mu\nu} \;,
        \]
and similar for $\tilde Z$, with the domain of integration $D_N$
defined by $\tilde Z^{\rm FF} = -{Z^{\rm FF}}^\dagger$, $\tilde Z^{\rm
  BB} = {Z^{\rm BB}}^\dagger$, and $1-{Z^{\rm BB}}^\dagger Z^{\rm BB}
> 0$.  To describe the integration measure, let $G_N = {\rm
  GL}(2N|2N)$ and $H_N = {\rm GL}(N|N)\times{\rm GL}(N|N)$, and define
$\phi_N$ to be the map that takes $Z,\tilde Z$ into the super coset
space $G_N/H_N$ by
        \[
        \phi_N(Z,\tilde Z) = \pmatrix{1 &Z\cr \tilde Z &1\cr} H_N \;.
        \]
Then $D\mu_N(Z,\tilde Z)$ arises by pulling back via $\phi_N$ the 
$G_N$-invariant Berezin measure on $G_N/H_N$.  It turns out that this
Berezin measure is locally flat:
        \[
        D\mu_N(Z,\tilde Z) = \prod_{\mu,\nu=1}^N
        {\rm d}Z_{\mu\nu}^{\rm BB} \wedge
        {\rm d}{\tilde Z}_{\mu\nu}^{\rm BB} \wedge
        {\rm d}Z_{\mu\nu}^{\rm FF} \wedge
        {\rm d}{\tilde Z}_{\mu\nu}^{\rm FF}
        {\partial^4 \over
        \partial Z_{\mu\nu}^{\rm BF} \partial{\tilde Z}_{\mu\nu}^{\rm BF}
        \partial Z_{\mu\nu}^{\rm FB} \partial{\tilde Z}_{\mu\nu}^{\rm FB}} 
        + ... \;.
        \]
(The dots indicate global anomalies.)  By differentiating the
supersymmetric integral for $\hat\Omega_U$ we obtain a formula for the
two-level correlation function:
        \begin{eqnarray}
       &&R_2(Nx/2\pi)={{\rm Re}\over 2N^2}\int_{D_N}D\mu_N(Z,\tilde Z)\,
        {\rm SDet}(1-\tilde ZZ)\,{\rm SDet}^{-1}(1-e^{ix}\tilde ZZ_U)
        \nonumber \\
        &&\hspace{2.8cm}
        \times \Big({\rm Tr} Z_U \tilde Z (e^{-ix} - Z_U\tilde Z)^{-1}
        \times {\rm Tr} \tilde ZZ_U (e^{-ix} - \tilde Z Z_U)^{-1}
        \nonumber \\
        &&\hspace{3.0cm}+\;e^{ix}{\rm Tr}Z_U\tilde Z(1-e^{ix}Z_U\tilde Z)
        ^{-1}\Sigma_3\tilde ZZ_U (1-e^{ix}\tilde ZZ_U)^{-1}\Big) \;,
        \nonumber
        \end{eqnarray}
where $Z_U = (1\otimes U)Z(1\otimes U^\dagger)$ and $\Sigma_3 = {\rm
  diag}(+1,-1) \otimes 1_N$.

While the steps done so far are exact and in fact rigorous, the
formula for $R_2(s)$ looks preposterous at first sight.  We have taken
a simple looking expression and transformed it into a horribly
complicated integral over a large supermatrix $Z$.  What's the
purpose?  The answer is that our formula promises to be a good
starting point for further analysis, at least for {\it chaotic} maps,
by the following reasoning.

\section{Zero Mode Approximation (heuristic)}

The goal is to make a statement about $\hat\Omega_U$ or $R_2$ in the
semiclassical limit.  Drawing the inspiration from recent work of
Andreev et al. \cite{a3s}, one might hope that for $\hbar \sim
N^{-1/d} \to 0$, the integral (\ref{Omega}) approaches a field
integral that will lend itself to evaluation by the methods of quantum
field theory.  However, as it stands, there is not much ground for
optimism.  Because there exists nothing that inhibits short wave
length fluctuations of the candidate field, the integral (\ref{Omega})
is certain {\it not} to converge to any well-defined field theory.
[Indeed, one can show that the integrand has $\sum_{n=0}^N
\pmatrix{N\cr n\cr}^2$ critical points, all of which contribute with
comparable weight to $R_2(s)$ for small $s$.  In the semiclassical
limit these critical points correspond to point-like instantons placed
on the periodic orbits of the map $\chi$, and the would-be field
theory looks like a dense gas of topological excitations.]

To make sense of the limit $N \to \infty$, one has no choice but to
introduce a {\it regulator}.  The natural procedure is to inject a
probabilistic component and do an ensemble average.  Thus, we will
abandon the ambition to treat an individual quantum map $U$ and,
instead, will average over a family of operators close to $U$, say
        \[
        U_\xi = e^{i\xi^k X_k / \hbar} U \;,
        \]
where $X_1, ..., X_r$ are Hermitian $N\times N$ matrices that arise 
from quantizing a set of Hamiltonian vector fields $\Xi_1, ...,\Xi_r$,
and the parameters $\xi^1,...,\xi^r$ are identically distributed
independent Gaussian random variables with zero mean and variance
$\epsilon$.

Substituting $U_\xi$ for $U$ in the expression for $\hat\Omega$ and
taking the expected value, we obtain
        \begin{eqnarray}
        &&E \left( \hat\Omega_{U_\xi}(\gamma_+;\gamma_-) \right)
        = \int D\mu_N(Z,\tilde Z) \; \exp(-S) \;, \nonumber \\
        &&S = - \ln E \Big( {\rm SDet}(1-\tilde ZZ) \; {\rm SDet}
        (1 - \tilde Z(\gamma_+ \otimes U_\xi)Z(\gamma_- \otimes
        U_\xi^\dagger))^{-1} \Big) \;. \nonumber
        \end{eqnarray}
The function $S$ has a critical point at $Z = \tilde Z = 0$.  For
$\gamma_+ = \gamma_- = 1_2$ this point extends to a supermanifold
$G_1/H_1$ of saddle points generated from $Z = \tilde Z = 0$ by the
$G_1$-action
        \begin{eqnarray}
        &&Z \mapsto g\cdot Z = (AZ+B)(CZ+D)^{-1} \;, \nonumber \\
        &&\tilde Z \mapsto g\cdot \tilde Z = 
        (C+D\tilde Z)(A+B\tilde Z)^{-1} \;, \nonumber
        \end{eqnarray}
where $A = a \otimes 1_N$, $B = b \otimes 1_N$, etc.  It is known
from \cite{a3s} that one gets the desired random matrix answer for
chaotic maps if one approximates the integral over $G_N/H_N$ by an
integral over the saddle point manifold $G_1/H_1$.  Hence, the natural
strategy is to try and justify this step, which is called the ``zero
mode approximation''.

The Hessian of $S$ at $Z = \tilde Z = 0$ (and, by $G_1$-symmetry,
everywhere else on the saddle point manifold) is given by the
following combination of superdeterminants:
        \[
        S^{(2)} = {\rm STr} \tilde Z Z - E \left( {\rm STr}
        \tilde Z U_\xi Z U_\xi^\dagger \right)
        \]
if $\gamma_+ = \gamma_- = 1_2$.  From this we see that the stability 
of the saddle point manifold is determined by the spectrum of $1 -
E(U_\xi \otimes U_\xi^\dagger)$.  Prior to ensemble averaging the
operator $1 - U_\xi \otimes U_\xi^\dagger$ has a large number $N$ of
zero modes (one for each eigenvector of $U_\xi$), and the
single-zero-mode approximation fails.  With ensemble averaging, the
situation is much improved.  One then expects $N-1$ zero modes to
become ``massive'' (leaving only a single massless mode), provided
that (i) the classical map $\chi$ is mixing, (ii) the squares of the
Hamiltonian vector fields $\Xi_1^2 + ... + \Xi_r^2$ add up to produce
an elliptic operator, and (iii) the parameter $\epsilon$ is taken to
vary with $\hbar$ as $\hbar^\alpha$ where $\alpha < 1$.  [This
conjecture comes from looking at the classical limit $1 - \exp(
\epsilon \sum_i \Xi_i^2) \circ \chi^*$ of the operator $1 - E( U_\xi
\otimes U_\xi^\dagger)$.  Moreover, from power counting one estimates
that, in order to achieve the desired mass gap, a positive value of
the exponent $\alpha$ suffices, which means that all members of the
ensemble converge to give the same classical dynamics in the limit
$\hbar\to 0$.]

Assuming the zero mode approximation to be justified, one arrives at
an integral over a $2 \times 2$ supermatrix $\zeta$:
        \[
        E\left(\hat\Omega_{U_\xi} (\gamma_+ ; \gamma_-)\right) \simeq
        \int_{D_1} D\mu_1(\zeta,\tilde\zeta) \,
        {\rm SDet}^N(1-\tilde\zeta\zeta) \,
        {\rm SDet}^{-N}(1-\tilde\zeta\gamma_+\zeta\gamma_-) \;.
        \]
This integral is independent of $U$ and is easy enough to be worked
out in closed form:
        \[
        \lim_{N\to\infty}
        E\Big(\hat\Omega_{U_\xi}(e^{ia_+/N},e^{ib_+/N};
        e^{-ia_-/N},e^{-ib_-/N})\Big)=
        1 - \frac{(a_+ - b_+)(a_- - b_-)}{(a_+ - a_-)
         (b_+ - b_-)} \left( 1-e^{i(b_+ - b_-)} \right) \;.
        \]
For the two-level correlation function one obtains $E(R_2(s)) = \delta
(s) - \sin^2(\pi s)/(\pi s)^2$, in agreement with the random-matrix
conjecture by Bohigas, Giannoni, and Schmit.

Clearly, there are many details that need to be filled in here.
Nonetheless, I believe the above approach to be promising and to have
the potential of leading to rigorous theorems.

{\bf Acknowledgment}. I would like to thank O. Agam and A. Altland for
discussions.  I especially thank S. Zelditch for making the suggestion
to compose the map $\chi$ with a stochastic Hamiltonian flow.

\end{document}